\documentclass[aip,amsmath,amssymb,reprint, graphicx]{revtex4-2}
\usepackage{graphicx}
\usepackage{xcolor}
\usepackage[export]{adjustbox}% http://ctan.org/pkg/adjustbox
\usepackage{hyperref}
\hypersetup{
    colorlinks=true,
    linkcolor=blue,
    citecolor=blue,
    filecolor=blue,
    urlcolor=blue
}
\begin{document}

\title{Electromagnetically induced transparency and population repump readout of Rydberg states of Cs atoms in a J-scheme} 

\author{Noah~Schlossberger}
\email[]{noah.schlossberger@nist.gov}
\affiliation{National Institute of Standards and Technology, Boulder, Colorado 80305, USA}
\author{Christopher~L.~Holloway}
\affiliation{National Institute of Standards and Technology, Boulder, Colorado 80305, USA}
\author{Erik~McKee}
\affiliation{TOPTICA Photonics, Pittsford, New York 14534, USA}
\author{Michael~A.~Highman}
\affiliation{TOPTICA Photonics, Pittsford, New York 14534, USA}
\author{Nikunjkumar~Prajapati}
\affiliation{National Institute of Standards and Technology, Boulder, Colorado 80305, USA}

\date{\today}

\begin{abstract}
Rydberg atom electrometry offers traceable electric field measurements over many decades of radio frequencies in a single device. Miniaturization of these sensors is primarily limited by requirements of the lasers used. Here we demonstrate a three-photon sensing scheme using a J-shaped energy level coupling that can be achieved using external cavity diode lasers, without the need for a doubling crystal or tapered amplifier. In the low laser power regime, we demonstrate a full-width at half-maximum linewidth of 1.3~MHz. We demonstrate that for RF field electrometry using conventional heterodyne techniques, we can detect 4.7~GHz at a sensitivity of 27~$\mu$V\,m$^{-1}$\,Hz$^{-1/2}$, comparable to that of two-photon detection schemes which require the use of a tapered amplifier. We also investigate a modified scheme where the probe laser is locked to a different hyperfine state, thus measuring the two-photon electromagnetically induced transparency in the other two lasers via the change in population of this separate state due to repumping. In this scheme we find the sensitivity for a 4.7~GHz field to be 39~$\mu$V\,m$^{-1}$\,Hz$^{-1/2}$, and demonstrate that the amplitude scaling with probe power offers a different saturation profile than the linked J-scheme counterpart.
\end{abstract}

%\pacs{}% insert suggested PACS numbers in braces on next line

\maketitle 

\section{Introduction}
Rydberg atoms have emerged as a powerful tool for SI-traceable, broadband electrometry \cite{Sedlacek2012,6910267,Jing2020,doi:10.1126/sciadv.ads0683,Schlossberger2024}. These sensors are laboratory-setting instruments, with deployability largely limited by the making of chip-scale lasers at the relevant wavelengths\cite{XING2025100187}. To reach the powers required at these wavelengths, frequency-doubling cavities and tapered amplifiers are often required\cite{PhysRevA.84.023408, PhysRevLett.98.113003, Hill:22}. Here, we present a three-photon readout scheme for which all three wavelengths can exist as standard external-cavity diode lasers without the need for these frequency doublers and tapered amplifiers. We use only a fiber amplifier for the near-infrared coupling laser, which can be compact and inexpensive. A similar energy level scheme was recently demonstrated\cite{ravi2025eitvinvertedxi} in Rb, and here we demonstrate and characterize the scheme in Cs. We characterize the linewidth of the resonant three-photon feature and its sensitivity for electrometry of radio-frequency fields. We also demonstrate a measurement scheme that utilizes a decay to couple a two-photon Rydberg ladder to the probe rather than a direct three-photon coherence. In this readout scheme, the observed spectral feature comes from population repumping of the ground state. We characterize the sensitivity of the repump readout and compare its linewidth and amplitude to its resonant three-photon counterpart.
\section{EIT readout}
\begin{figure}
\includegraphics[scale = .9]{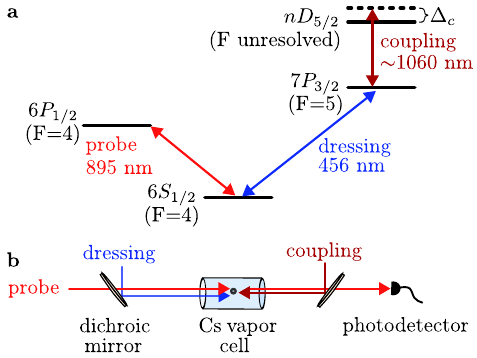}
\caption{The ``J''-shaped electromagnetically induced transparency scheme. \textbf{a} The energy level diagram of the measurement scheme. \textbf{b} The configuration of the lasers' $k$-vectors.}
\label{fig:scheme}
\end{figure}

The scheme utilizes a ``J''-shaped chain of lasers to probe the energy of the Rydberg state. The energy level diagram is shown in Fig. \ref{fig:scheme}a.
The direction of the beams (Fig. \ref{fig:scheme}b) is chosen to minimize the the Doppler k-vector mismatch, given by
\begin{multline}
|\sum_i s_i k_i| =\\
 \left|
(-1)\left(\frac{2\pi}{895 \textrm{ nm}}\right) + 
(1)\left(\frac{2\pi}{465 \textrm{ nm}}\right) + 
(1)\left(-\frac{2\pi}{1060 \textrm{ nm}}\right)
\right|\\
= 0.6~\mu\text{m}^{-1}
\end{multline}
where $k_i$ is the signed $k$-vector along the axis of the beams.
The parameter $s_i$ is either $\pm1$, where the plus sign indicates that the energy has increased going from one state to the other and a minus sign indicates the energy has decreased going from one state to the other. In our case, we start at 6P$_{1/2}$, $s_i=-1$ going from 6P$_{1/2}$ to 6S$_{1/2}$ (the probe line), and $s_i=1$ for both the dressing and coupling lines. This residual $k$-vector is a factor of eight lower than that of a counter-propagating 852~nm and 510~nm two-photon scheme\cite{Zhao:09} which has a residual $k$-vector of 4.9 $\mu$m$^{-1}$.

%where $s_i$ is the sign of the energy difference across each successive coupling of %the J scheme and $k_i$ is the signed $k$-vector along the axis of the beams. This %residual $k$-vector is a factor of eight lower than that of a counter-propagating %852~nm and 510~nm two-photon scheme\cite{Zhao:09} which has a residual $k$-vector of %4.9 $\mu$m$^{-1}$.

The probe frequency is locked to a saturated absorption spectrum in a reference cell, the dressing laser frequency is locked to a two-photon co-propagating EIT spectrum between the probe and dressing in a reference cell, and the coupling laser frequency is scanned to read out the Rydberg state. The transmission of the probe beam is read out on a photodetector.
\subsection{Linewidth}
\begin{figure}[h]
\includegraphics[scale = .9]{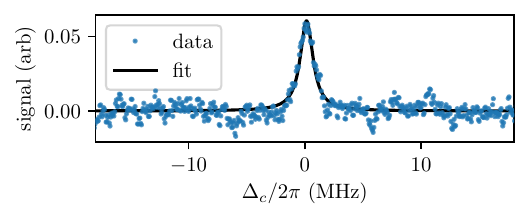}
\caption{Lineshape of the three-photon transparency on the $50D_{5/2}$ state in the low-power regime.}
\label{fig:narrowline}
\end{figure}

To optimize the linewidth, the laser Rabi rates must be reduced such that they are not the dominating broadening effect. The parameters of the lasers are given in Table \ref{tab:laserparamsnarrow}.
\begin{table}[h]
\begin{tabular}{l|l |l |l}
    laser & power ($\mu$W) & FWHM (mm) & $\Omega$ ($2\pi \times\text{MHz}$)\\
    \hline
    probe & 1.3 & 1.6 & $2\pi \times 0.72$\\
    dressing & 17 & 2.2 & $2\pi \times 0.26$\\
    coupling & 460 & 2.4 & $2\pi \times 0.17$
\end{tabular}
\caption{Laser parameters used for narrow EIT linewidth.}
\label{tab:laserparamsnarrow}
\end{table}\\
To compensate for the correspondingly small signal at these low laser powers, we use an optical chopper wheel to modulate the dressing light and perform lock-in detection of the probe transmission to achieve adequate signal-to-noise ratio (SNR). We also wrap the vapor cell in high permeability magnetic shielding, as background magnetic fields can cause spectral broadening on the order of MHz.\cite{PhysRevA.109.L021702} With this, we scan the coupling laser detuning $\Delta_c$ from its resonance and monitor the change in probe transmission. The frequency axis scaling of the coupling detuning is calibrated with an ultra-stable cavity by referencing the spacing of cavity resonances between two electro-optic modulator sidebands that differ in frequency by 1.00 MHz. The resulting spectral lineshape is shown in Fig. \ref{fig:narrowline}.
 
We fit the lineshape to a Lorentzian with a full-width at half-maximum (FWHM) linewidth of $1.32 \pm 0.06$ MHz. This linewidth is still likely power-broadening-limited, but further reduction in laser powers is limited by SNR constraints. However, it is still well below the two-photon linewidth limit for energy resolution of the Rydberg state \cite{schlossberger2026fundamentallinewidthlimitelectromagnetically}.
\subsection{Electrometry and sensitivity}
\begin{figure}[h]
\textbf{a}\includegraphics[scale = .9, valign = t]{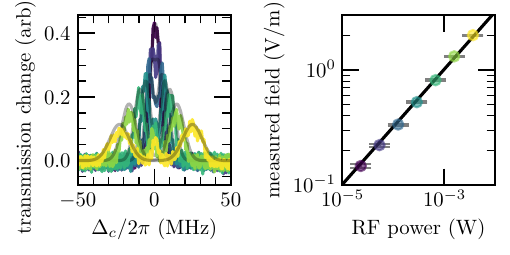}\\
\textbf{b}\includegraphics[scale = .9, valign = t]{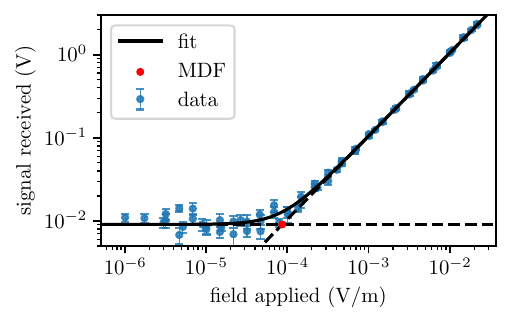}
\caption{Sensitivity measurement of the system on the 4.7 GHz $53D_{5/2} \rightarrow 54P_{3/2}$ transition. \textbf{a} The Autler-Townes splitting of the $53D_{5/2}$ state is measured at various powers (left), and the field inferred from the splitting is used to calibrate the field at the atoms as a function of the power applied to the horn (right). \textbf{b} With a local oscillator, an 11~kHz beatnote is formed, and the minimum detectable field (MDF) is determined for a resolution bandwidth of 10~Hz.}
\label{fig:sens}
\end{figure}

We now consider this technique as a method for electrometer of radio-frequency (RF) fields. We remove the magnetic shielding to allow RF to enter the cell. We detect the RF field with a heterodyne detection scheme as in Refs.\cite{10.1063/1.5095633,Jing2020}. While low power regime shown in the previous section gives narrow lines, it gives poor SNR. The sensitivity, proportional to the slope of the spectral feature, is optimized for higher laser powers. The height of the feature can be improved by over an order of magnitude in exchange for a factor of a few in linewidth, meaning the slope can be enhanced with more laser power. We find the optimized regime to be powers of 18~$\mu$W, 380~$\mu$W and 419~mW for the probe, dressing, and coupling lasers, resulting in Rabi rates of $2\pi\times 2.7$~MHz, $2\pi\times 1.2$~MHz, and $2\pi\times 5.13$~MHz respectively. At these powers, we measure the sensitivity of the $53D_{5/2}\rightarrow 54P_{3/2}$ transition in Fig.~\ref{fig:sens}.

First, we calibrate the strength of the field we apply by looking at Autler-Townes splitting to determine the relationship between the root of the power we apply to the microwave horn and the electric field at the location of the atoms  (Fig. \ref{fig:sens}a). Then, we apply an RF local oscillator and measure the beatnote with a spectrum analyzer, and decrease the field we apply until we run into the noise floor  (Fig. \ref{fig:sens}b). See Ref.\cite{10.1063/1.5095633} for more details on this approach. We fit the signal $V_\textrm{S.A.}$ received on the spectrum analyzer in the relevant frequency band to:
\begin{equation}
V_\textrm{S.A.} = \sqrt{V_\textrm{N.E.}^2 + (A\cdot E_\textrm{RF})^2},
\end{equation}
where $V_\textrm{N.E.}$ is the noise-equivalent voltage on the spectrum analyzer (due to all system noises), $E_\textrm{RF}$ is the magnitude of the applied RF field, and $A$ is some dimensionful constant capturing the conversion gain of the detection system. We float $V_\textrm{N.E.}$ and $A$. The minimum detectable field $E_\textrm{MDF}$, that is the field at which the signal-to-noise ratio (SNR) is exactly 1, is given by 
\begin{equation}
E_\textrm{MDF} = V_\textrm{N.E.}/A.
\end{equation}
We then find the sensitivity $\mathcal{S}$ by dividing by the root of the resolution bandwidth (RBW):
\begin{equation}
\mathcal{S} = \frac{E_\textrm{MDF}}{\sqrt{\textrm{RBW}}}.
\end{equation}
Here, the RBW is equivalent to the inverse of the integration time of the measurement. With a minimum detectable field of $86\pm3$ ~$\mu$V/m at an RBW of 10~Hz, we find the sensitivity to be $27\pm 1$~$\mu$V\,m$^{-1}$\,Hz$^{-1/2}$. This sensitivity is comparable to that of similar electrometry techniques in two-photon EIT schemes \cite{Jing2020, jsbl-45t9, 10.1063/5.0086357,6910267,Sedlacek2012,10.1063/1.5095633}. 
\section{Repump readout}
\begin{figure}[h]
\includegraphics[scale = .9]{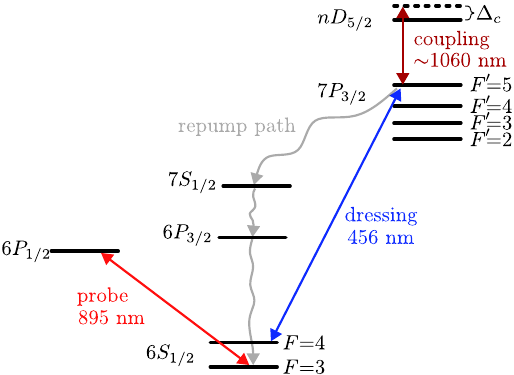}
\caption{Energy level diagram of the repump readout scheme.}
\label{fig:repumpscheme}
\end{figure}

In addition to the readout scheme in Fig. \ref{fig:scheme}a, we can employ a similar scheme in which the probe and dressing do not address the same hyperfine state. This scheme requires a natural decay in order for the Rydberg state dynamics to influence absorption of the probe. Such a scheme is shown in Fig. \ref{fig:repumpscheme}.

Here, the probe transmission essentially represents the  population in the $6S_{1/2}, (F=3)$ state (within the Doppler velocity class in which the three lasers are resonant). While no other lasers talk coherently to this state, its population is effected by decay from the $7P_{3/2}$ state, so population in this state enhances the absorption through population repumping. With the dressing locked on resonance, population in $F=4$ is repumped into $F=3$, thus increasing absorption of the probe. However, when the coupling is on resonance, a two-photon EIT condition will occur, trapping population in a dark state that does not repump the $F=3$ population. As such, when the coupling is on resonance there is a decrease in population repumping and a decrease in probe absorption.

\begin{figure}[h]
\includegraphics[scale = .9]{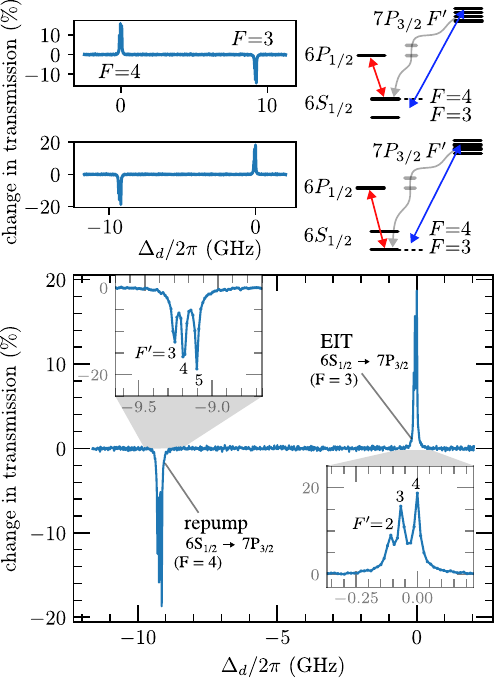}
\caption{Two-photon (probe, dressing) resonance scan displaying both the EIT and repump features with the probe locked to the $F=4$ state (top) and the $F=3$ state (middle, bottom). On the bottom plot, the $F$ of the $6S_{1/2}$ state and the $F'$ of the $7P_{3/2}$ state are labeled for each spectral feature.
}
\label{fig:lambdarepump}
\end{figure}

First, we observe the effect of population repumping without the coupling laser, scanning the dressing laser instead. In this two-photon $\lambda$-scheme EIT between the probe and the dressing, we send the beams copropagating to optimally restrict the Doppler velocity classes. The spectra are shown in Fig. \ref{fig:lambdarepump}.

Here we define $\Delta_d$ as the detuning of the dressing light from the $6S_{1/2} \rightarrow 7P_{3/2}$ transition that addresses the same hyperfine state as the probe. We find that the repump features are actually narrower than the EIT features because the Doppler selection criteria is the same (provided that the collision rate is low compared to the detection photon cycling rate), but the criteria for the probe power to saturate is more relaxed (the $6P_{1/2}$ state decays faster than the dark state of the EIT feature).

\begin{figure}
    \includegraphics[scale=.9]{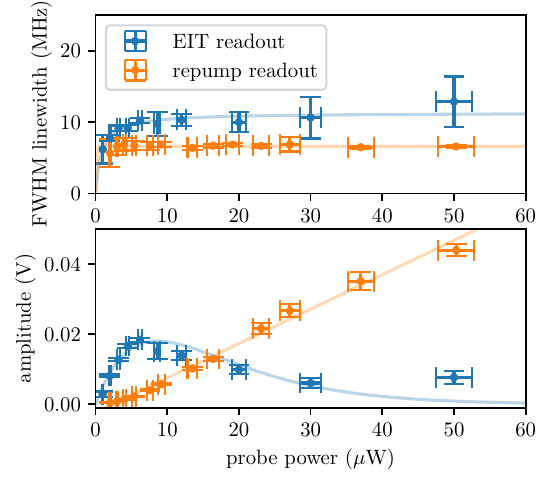}
    \caption{Comparison of the linewidth and relative amplitude of EIT and repump spectra of the $53D_{5/2}$ state for various probe powers and with a dressing power of 0.6~mW ($\Omega_d = 2\pi \times 1.5$~MHz) and a coupling power of 26~mW ($\Omega_c = 2\pi\times 1.28$~MHz).}
    \label{fig:amplinewidth}
\end{figure}

Locking the probe laser to $F=3$ and the dressing laser to $F=4$, we can introduce the coupling laser to enable repump readout of the Rydberg state. We use the same experimental layout to that in Fig. \ref{fig:scheme}b, with the only practical difference being the frequency of the probe laser. Features of this repump readout (probe locked to $F=3$) method are compared to that of the three-photon EIT (probe locked to $F=4$) method in Fig. \ref{fig:amplinewidth}.
Once again, we find the repump method to be slightly narrower in linewidth. Furthermore, we find that the amplitude of the signal saturates with increased probe power for the EIT readout, but scales linearly in power for the repump readout. 

\begin{figure}
\includegraphics[scale = .9]{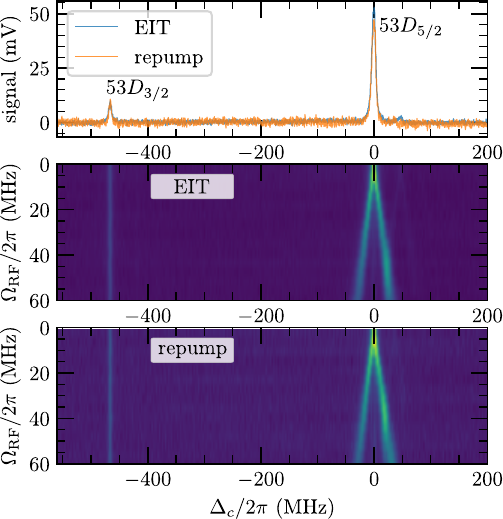}
\caption{EIT and repump readout of the $53D$ states. The lineshapes are compared for the same laser powers (top), and the Autler-Townes spectra are compared as waterfall plots when coupling the $53D_{5/2} \rightarrow 54P_{3/2}$ states (middle, bottom).}
\label{fig:EITrepumplineshape}
\end{figure}

Next, we compare the two readout methods in the context of electrometry. We measure the Autler-Townes spectra of the 4.7~GHz $53D_{5/2} \rightarrow 54P_{3/2}$ transition in Fig. \ref{fig:EITrepumplineshape}.
We find both methods are capable of field electrometry, with better SNR from the EIT readout method. We perform an analysis identical to that of Figure \ref{fig:sens} with the repump readout and find the sensitivity to be 39$\pm$1~$\mu$V\,m$^{-1}$\,Hz$^{-1/2}$, slightly poorer than that of the EIT readout scheme.

Even though the repump method allows for increased probe power without spectral broadening, this does not necessarily translate to higher sensitivity in the context of RF electrometry. The worse sensitivity of the repump readout method is likely because of two factors:
\begin{itemize}
    \item Only some of the decay from the $7P_{3/2}$ state goes into each $F$ level of the $6S_{1/2}$ state, meaning some of the RF-sensitive population change is not observed.
    \item Only some of the population in the $F$ state that absorbs probe light comes from this decay, meaning that much of the signal represents background.
\end{itemize}
\section{Conclusion}
We have demonstrated a three-photon sensing scheme for the Rydberg states of Cs using only external cavity diode lasers. We have shown that for low powers, we can achieve linewidths on the order of one MHz and field sensitivities as low as 27~$\mu$V\,m$^{-1}$\,Hz$^{-1/2}$, comparable with two-photon EIT schemes.  This technique offers several advantages over conventional two-photon EIT readout schemes. The lasers used do not require a frequency doubling crystal or a tapered amplifier. In addition, high power visible light can cause line-broadening charge distributions in the vapor cell due to the photoelectric effect on surface alkali atoms\cite{10.1116/5.0264378}. Here, the required visible light power is much less because it addresses a transition with a much larger dipole moment. 
We have also characterized a repump readout method using the same lasers, and shown a similar sensitivity. In the future, investigation into fluorescence-based readout schemes using these energy level couplings could offer sensitivity improvements by removing noise associated with background photon counts from the measured signal\cite{10.1116/5.0201928}.

\begin{acknowledgments}
\noindent
The authors thank William~J.~Watterson and Ying-Ju Wang for constructive comments during the internal review of this manuscript.\\
This research was supported by NIST under the NIST-on-a-Chip program.  A contribution of the U.S. government, this work is not subject to copyright in the United States. Any collaboration associated with the work does not imply recommendation or endorsement of any product or service by NIST, nor is it intended to imply that any such products or services are necessarily the best available for the purpose.\\
Data associated with this publication is publicly available. \cite{MIDAS}\\
The authors declare no conflict of interest.
\end{acknowledgments}

%\bibliography{main}
%aipnum4-2.bst 2019-01-14 (MD) hand-edited version of apsrev4-1.bst

\end{document}